\documentclass[12pt]{article}
\usepackage{amsmath}
\usepackage{graphicx,psfrag,epsf}
\usepackage{enumerate}
\usepackage{natbib}
\usepackage[hyphens]{url} 

\usepackage{bm} 
\usepackage{ams fonts} 
\usepackage{booktabs}
\usepackage{pdflscape} 
\usepackage{multirow} 

\newcommand{\blind}{1}

\begin{document}

\def\spacingset#1{\renewcommand{\baselinestretch}%
{#1}\small\normalsize} \spacingset{1}


\if1\blind
{
  \title{\bf Incorporating Prior Information with Fused Sparse Group Lasso: Application to Prediction of Clinical Measures from Neuroimages}
  \author{Joanne C. Beer \thanks{
    Joanne C. Beer is a Doctoral Candidate in Biostatistics at the Department of Biostatistics, University of Pittsburgh, Pittsburgh, PA 15261 (e-mail: joannecbeer@pitt.edu). Howard J. Aizenstein is a Professor of Psychiatry (primary appointment) and Bioengineering and Clinical and Translational Science (secondary appointment) at the Department of Psychiatry, University of Pittsburgh School of Medicine, Pittsburgh, PA, 15213 (e-mail: aizensteinhj@upmc.edu). Stewart J. Anderson is a Professor of Biostatistics (primary appointment) and Clinical and Translational Science (secondary appointment) at the Department of Biostatistics, University of Pittsburgh, Pittsburgh, PA 15261 (e-mail: sja@pitt.edu). Robert T. Krafty is an Associate Professor of Biostatistics at the Department of Biostatistics, University of Pittsburgh, Pittsburgh, PA 15261 (e-mail: rkrafty@pitt.edu). This work was partially supported by the National Institute of General Medical Sciences grant R01GM113243 and National Institute on Aging grants 2P01AG025204-11A and R01AG052446. The authors thank George C. Tseng, Helmet Karim, and Dana Tudorascu for helpful comments and suggestions. } \hspace{.2cm}\\
    Department of Biostatistics, University of Pittsburgh\\
    Howard J. Aizenstein \\
    Department of Psychiatry, University of Pittsburgh\\
    Stewart J. Anderson \\
    Department of Biostatistics, University of Pittsburgh\\
    and \\
    Robert T. Krafty \\
    Department of Biostatistics, University of Pittsburgh}
  \maketitle \newpage
} \fi

\if0\blind
{
  \bigskip
  \bigskip
  \bigskip
  \begin{center}
    {\LARGE\bf Incorporating Prior Information with Fused Sparse Group Lasso: \\ \vspace{0.25mm} Application to Prediction of Clinical \\ \vspace{1.5mm} Measures from Neuroimages}
\end{center}
  \medskip
} \fi

\bigskip
\begin{abstract}
Predicting clinical variables from whole-brain neuroimages is a high dimensional problem that requires some type of feature selection or extraction. Penalized regression is a popular embedded feature selection method for high dimensional data. For neuroimaging applications, spatial regularization using the $\ell_1$ or $\ell_2$ norm of the image gradient has shown good performance, yielding smooth solutions in spatially contiguous brain regions. However, recently enormous resources have been devoted to establishing structural and functional brain connectivity networks that can be used to define spatially distributed yet related groups of voxels. We propose using the fused sparse group lasso penalty to encourage structured, sparse, interpretable solutions by incorporating prior information about spatial and group structure among voxels. We present optimization steps for fused sparse group lasso penalized regression using the alternating direction method of multipliers algorithm. With simulation studies and in application to real fMRI data from the Autism Brain Imaging Data Exchange, we demonstrate conditions under which fusion and group penalty terms together outperform either of them alone. Supplementary materials for this article are available online.
\end{abstract}

\noindent%
\vfill

\newpage
\spacingset{1.45} 
\section{Introduction}
\label{sec:intro}
Despite nearly 30 years of functional neuroimaging research, there have been relatively few translations of basic neuroscience findings to clinical applications in psychiatry, such as the use of biomarkers for determining diagnosis, prognosis, or predicting treatment response \citep{kapur2012has,woo2017building}. The traditional mass univariate approach in neuroimaging, which fits a model to each voxel independently, has been successful at characterizing group-level brain structure and function. However, a predictive model approach, where neuroimage features serve as predictors and a clinical variable is modeled as the outcome, may be better suited to clinical application (e.g.,\ \cite{wager2013fmri}). Predictive models are able to exploit dependencies between brain regions and thus can potentially explain more variability in the outcome than a mass univariate approach. Moreover, predictive models allow inference at the individual level such that a prediction can be obtained for a new individual whose data was not used to train the model.

In this paper, we show how the fused sparse group lasso, a structured, sparse estimator, can incorporate prior information into a predictive model, thereby allowing researchers to harness results from the extensive recent research on brain structural and functional connectivity. Our goals include not only predictive accuracy, gauged by how well the model predicts the response for independent test data, but also interpretable parameter estimates, as based on the following criteria: (1) Model structure entails that parameter values have a straightforward meaning; e.g.,\ linear models tend to be more interpretable than nonlinear models.  (2) Models are appropriately sparse, including only relevant predictors, while not excluding any relevant predictors. (3) Parameter estimates are understandable in light of existing background knowledge. In a translational neuroimaging context, this would mean that the brain regions implicated by the model estimates are neuroscientifically plausible according to existing knowledge or provide new insight into the neurobiological mechanism influencing the clinical outcome, and can potentially be used to establish biomarkers. 

In Section \ref{sec:app}, we apply fused sparse group lasso to a resting state functional magnetic resonance imaging (fMRI) dataset from the Autism Brain Imaging Data Exchange, a public repository of MRI datasets \citep{di2014autism}. Autism spectrum disorder (ASD) is a group of developmental disorders characterized by impaired social functioning and restrictive and repetitive behavior, and affects approximately 1\% of children \citep{baio2012prevalence}. Neuroimaging studies report abnormal functional connectivity between brain regions in ASD, although findings are mixed regarding the specific nature of the abnormalities \citep{di2014autism}. In our application, we show that incorporating prior information about voxel spatial location and functional connectivity using fused sparse group lasso increases accuracy when predicting a continuous measure of autistic social impairment from resting state fMRI data.

While interest in predicting continuous outcomes is increasing, the majority of work in prediction from neuroimages has focused on classification problems \citep{cohen2011decoding,arbabshirani2017single}. Numerous studies have built classifiers to differentiate patients from healthy controls or sort patients into diagnostic groups on the basis of neuroimaging data. However, many psychiatric diagnostic categories are of questionable validity and may group together individuals with heterogenous etiology underlying their symptoms \citep{insel2010research}. Continuous measures of more fundamental phenotypic traits may map better onto underlying neurobiology and may be particularly suitable for spectrum disorders such as autism. While our proposed penalty could easily be extended for use in conjunction with a variety of loss functions, including those for categorical outcomes such as logistic loss, here we adopt a dimensional approach using continuous outcomes. 

Consider the linear regression model
\begin{equation} \label{eq:linreg}
\mathbf{y} = \mathbf{X} \bm{\beta} + \bm{\epsilon},
\end{equation}
where $\mathbf{y} \in \mathbb{R}^n$ is a continuous outcome (e.g.,\ depression score), 
$\mathbf{X} \in \mathbb{R}^{n \times p}$ is a predictor matrix (e.g.,\ neuroimage voxel values),
$\bm{\beta} \in \mathbb{R}^p$  is an unknown vector of coefficients,
$\bm{\epsilon}  \in \mathbb{R}^n$ is the error, $E(\bm{\epsilon}) = \bm{0}$, 
and $E(\bm{\epsilon}^T \bm{\epsilon}) = \sigma^2 \mathbf{I}_n$.
This represents a high dimensional setting where the number of subjects $n$ is much less than the number of predictors $p$, which can be on the order of 100,000 voxels. To obtain a unique solution for $\bm{\beta}$, we can constrain the optimization problem using the penalized least squares estimator
\begin{equation} \label{eq:estimator}
\widehat{\bm{\beta}} = {\arg \min}_{\bm{\beta} \in \mathbb{R}^p} \frac{1}{2} \| \mathbf{y} - \mathbf{X} \bm{\beta} \|_2^2 + \lambda  \, J( \bm{\beta} ),
\end{equation}
where $\lambda \ge 0$ is a tuning parameter controlling the level of regularization. The penalty term $J( \bm{\beta} )$ can impose both sparsity and structure, thereby constraining the solution space according to \textit{a priori} information about relationships between elements of $ \bm{\beta} $. In a neuroimaging context, for example, this information might include spatial proximity or previously established functional connectivity networks.  Examples of unstructured and structured penalties are presented in Table \ref{tab:penalties}.

\begin{table}
\small
\caption{Examples of unstructured and structured penalty terms  
\label{tab:penalties}}
\begin{center}
\footnotesize
\begin{tabular}{ll}
\hline
Unstructured penalties								&  $J( \bm{\beta} )$                  									\\
\hline
Lasso  \citep{tibshirani1996regression}                             		& $\| \bm{\beta} \|_1$                          							\\
Ridge   \citep{hoerl1970ridge}							& $\| \bm{\beta} \|_2^2$     									\\
Elastic net   \citep{zou2005regularization}        				& $\alpha \| \bm{\beta} \|_1 + (1 - \alpha)  \| \bm{\beta} \|_2^2; \ \alpha \in [0,1]$	\\
\hline
Structured penalties            							&  $J( \bm{\beta} )$		   	\\
\hline
Isotropic total variation \citep{rudin1992nonlinear}			&  $\| \mathbf{D} \bm{\beta} \|_{2,1}$; matrix $\mathbf{D}$ encodes spatial structure   \\
Fused lasso*  \citep{tibshirani2005sparsity}				& $\alpha \| \bm{\beta} \|_1 + (1 - \alpha) \| \mathbf{D} \bm{\beta} \|_1; \ \alpha \in [0,1]$ \\
Graph net**  \citep{grosenick2013interpretable}				& $\alpha \| \bm{\beta} \|_1 + (1 - \alpha) \| \mathbf{D} \bm{\beta} \|_2^2; \ \alpha \in [0,1]$\\
Group lasso  \citep{yuan2006model}						& $\sum_{g \in \mathcal{G}} \sqrt{p_g} \| \bm{\beta}_g \|_2$; groups $\mathcal{G}$ form a partition of $\bm{\beta}$ \\
Sparse group lasso 	\citep{simon2013sparse}				& $\alpha \| \bm{\beta} \|_1 + (1 - \alpha) \sum_{g \in \mathcal{G}} \sqrt{p_g} \| \bm{\beta}_g \|_2; \ \alpha \in [0,1]$                          	\\
\hline
*Also known as anisotropic total variation--$\ell_1$					& \\
**Also known as sparse graph Laplacian 						&
\end{tabular} 
\end{center}
\end{table}

Simulation studies and applications to real neuroimaging datasets (mostly using functional magnetic resonance imaging (fMRI)) have shown that penalties enforcing spatial smoothness frequently outperform unstructured penalties \citep{michel2011total,baldassarre2012structured,gramfort2013identifying,grosenick2013interpretable,fiot2014longitudinal,xin2014efficient}. Not only do spatially-informed penalties yield more interpretable estimates insofar as they select contiguous groups of voxels in neuroscientifically plausible brain regions, but they often show better prediction performance. For example, \cite{michel2011total} found that the isotropic total variation penalty gave higher prediction accuracy and recovered the support of the true parameters better than elastic net and linear support vector regression in simulation studies, and for real fMRI data produced estimates that were less dispersed and more neuroscientifically interpretable. \cite{baldassarre2012structured} reported that the anisotropic total variation-$\ell_1$ penalty (equivalent to a three dimensional fused lasso) had the highest classification accuracy in a real fMRI data application as compared with lasso, elastic net, graph net, and anisotropic total variation alone. Additionally, it produced the most stable estimates across validation folds. Both of these studies involved healthy participants viewing various images in the fMRI scanner, and models were designed to predict or classify some feature of the images. In a more clinically-oriented study, \cite{fiot2014longitudinal} used structural neuroimaging data to predict Alzheimer's disease progression. The authors compared several penalties, including ridge, lasso, elastic net, non-sparse graph net, isotropic total variation, and isotropic total variation-$\ell_1$. The spatially-informed penalties yielded more neuroscientifically relevant coefficient maps and statistically better classification accuracy than the unstructured penalties.

In addition to spatial regularization, group-structured regularization has shown promise in predictive neuroimaging models. \cite{shimizu2015toward} compared logistic regression using lasso, group lasso, and sparse group lasso penalties, linear support vector machine (SVM), and random forest for classifying depression patients and healthy controls based on fMRI data. For the group lasso penalties, voxels were grouped according to known functional and anatomical brain regions. The authors found that group lasso and sparse group lasso were superior to lasso and random forest and comparable to SVM in terms of classification accuracy, but unlike SVM, they produced sparse and more interpretable models. Rather than defining voxel groups \textit{a priori}, \cite{liu2014identifying} used a data-driven agglomerative hierarchical clustering method to create a tree-structured grouping of voxels in grey matter density brain maps. Feature selection was then performed using a tree-structured group lasso penalty, and the selected features were used in a linear SVM to discriminate Alzheimer's disease patients from healthy controls. The proposed method achieved higher classification accuracy than feature selection using the $\ell_1$ lasso penalty or an anatomically-defined group lasso penalty.

For neuroimaging applications, we aim to incorporate two types of structure into the penalty term $J(\bm{\beta})$ of the estimator in Equation (\ref{eq:estimator}): (1) local spatial information, to encourage smooth coefficient estimates across neighboring voxels; and (2) spatially distributed groups, such as those defined by functional or structural networks or anatomical regions, to allow voxels within the same group to be selected or shrunk to zero together.  We achieve this by combining $\ell_1$, fusion, and group lasso penalties into a fused sparse group lasso penalty. We found one instance of this penalty in the literature, in a multi-task learning context where groups consist of repeated measures of the same task and smoothing is applied across time points within a group \citep{zhou2012modeling}. To our knowledge, the fused sparse group lasso penalty has never been studied via simulations or used in a predictive model with neuroimaging data. 

In the remainder of this paper, we present the fused sparse group lasso estimator in Section \ref{sec:fsgl} and derive update steps to fit the fused sparse group lasso penalized least squares regression model using the alternating direction method of multipliers algorithm \citep{boyd2011distributed} in Section \ref{sec:opt}. We report methods and results of a simulation study in Section \ref{sec:sim} and apply our method to resting state fMRI data from the Autism Brain Imaging Data Exchange repository in Section \ref{sec:app}. We make concluding remarks in Section \ref{sec:conc}. We provide R and MATLAB functions for fitting the fused sparse group lasso estimator and additional supplementary materials online.

\section{Fused Sparse Group Lasso}
\label{sec:fsgl}

\subsection{Model}
\label{sec:fsglmodel}
Suppose we observe $ \{(\mathbf{x}_1, y_1), \ldots, (\mathbf{x}_n, y_n) \}$ from $n$ independent subjects, indexed by $i = 1, \ldots, n$, where $y_i \in \mathbb{R}$ and $\mathbf{x}_i \in \mathbb{R}^p$. In the neuroimaging context considered here, $y_i$ is a continuous scalar outcome for each subject such as age, depression score, or cognitive test score, and $\mathbf{x}_i$ is a vector of voxel values from a three dimensional brain image such that each element of $\mathbf{x}_i$ corresponds to one of $p$ voxels. Assume that $\mathbf{y} = (y_1, \ldots, y_n)^T$ and the columns of the matrix $\mathbf{X} = (\mathbf{x}_1 | \ldots | \mathbf{x}_n)$ are centered, so we do not have an intercept term. Furthermore, we standardize the columns of $\mathbf{X}$ to have standard deviation of one. We model the continuous outcome using standard linear regression as expressed in Equation (\ref{eq:linreg}).

\subsection{Estimator}
Since the number of voxels is typically orders of magnitude larger than the number of subjects, i.e.,\ $p \gg n$, regularization is required to obtain a unique solution for $\bm{\beta}$. We propose estimating $\bm{\beta}$ by minimizing the sum of the loss function and three penalty terms:
\begin{equation} \label{eq:opt}
\hat{\bm{\beta}} = {\arg \min}_{\bm{\beta} \in \mathbb{R}^p}   \ L(\bm{\beta}) + \lambda_1 \| \bm{\beta} \|_1 + \lambda_2 \| \mathbf{D} \bm{\beta} \|_1 + \lambda_3   \Omega^{\mathcal{G}} (\bm{\beta}); 
\end{equation}
where $L(\bm{\beta})$ is the loss function (e.g.,\ least squares); 
$\| \bm{\beta} \|_1 = \sum_{j=1}^p \lvert \beta_j \rvert $ is the $\ell_1$ norm of $\bm{\beta}$;
$\mathbf{D}_{m \times p}$ is the three dimensional fusion matrix for fused lasso, e.g., for a $2 \times 2 \times 2$ cubic image,
\begin{equation*}
\mathbf{D} = 
\left[
\begin{array}{cccccccc}
1 & 0  & -1 & 0 & 0 & 0 & 0 & 0 \\
0 & 1  & 0 & -1 & 0 & 0 & 0 & 0 \\
0 & 0  & 0 & 0 & 1 & 0 & -1 & 0 \\
0 & 0  & 0 & 0 & 0 & 1 & 0 & -1 \\

1 & -1  & 0 & 0 & 0 & 0 & 0 & 0 \\
0 & 0  & 1 & -1 & 0 & 0 & 0 & 0 \\
0 & 0  & 0 & 0 & 1 & -1 & 0 & 0 \\
0 & 0  & 0 & 0 & 0 & 0 & 1 & -1 \\

1 & 0 & 0 & 0 & -1  & 0 & 0 & 0 \\
0 & 1 & 0 & 0 & 0  & -1 & 0 & 0 \\
0 & 0 & 1 & 0 & 0  & 0 & -1 & 0 \\
0 & 0 & 0 & 1 & 0  & 0 & 0 & -1 \\
\end{array}
\right],
\end{equation*} 
and $\| \mathbf{D} \bm{\beta} \|_1$ is the fusion penalty; 
$ \Omega^ {\mathcal{G}} (\bm{\beta})  = \sum_{g \in \mathcal{G}} \sqrt{p_g} \| \bm{\beta}_g \|_2 $ is the $\ell_{2,1}$ group lasso penalty, which applies the $\ell_{2}$ norm, $\| \bm{\beta}_g \|_2 = \sqrt{ \bm{\beta}_g^T \bm{\beta}_g }$, to the coefficients $\bm{\beta}_g$ for each group $g \in \mathcal{G}$, each of size $p_g$;
and $\lambda_1, \lambda_2, \lambda_3 \ge 0$ are regularization tuning parameters. 

The three penalty terms incorporate prior information into the estimator, encouraging the solution to have both sparsity and a particular structure. The standard lasso $\ell_1$ penalty encourages overall sparsity. The fusion penalty penalizes the absolute differences between coefficients at neighboring voxels, thereby encouraging local smoothness. The group lasso penalty encourages a group-level structure; entire groups may be selected or shrunk to zero together. For example, if groups are defined by functional networks, the penalty allows voxels involved in a common network to be shrunk to zero if that network is not important for prediction. Given the overlapping structure of brain networks, overlapping groups are another possibility worth considering. With appropriate weighting and a latent variable approach \citep{jacob2009group,obozinski2011group}, the estimator could also accommodate overlapping groups. However, that is beyond the scope of the current paper. 

For ease of selecting values for the tuning parameters via cross-validation, it is convenient to reparameterize (\ref{eq:opt})  as follows:
\begin{equation} \label{eq:opt2}
\hat{\bm{\beta}} = {\arg \min}_{\bm{\beta} \in \mathbb{R}^p}   \ L(\bm{\beta}) + \alpha \gamma \lambda \, \| \bm{\beta} \|_1 + (1 - \gamma) \lambda \, \| \mathbf{D} \bm{\beta} \|_1 + (1 - \alpha) \gamma \lambda  \, \Omega^{\mathcal{G}} (\bm{\beta}), 
\end{equation}
such that $\lambda > 0$ controls the overall level of regularization, $\alpha \in [0, 1]$ controls the balance between the two sparsity inducing penalties (lasso and group lasso), and $\gamma \in [0, 1]$ controls the balance between the two sparsity inducing penalties and the fusion penalty. When $\alpha = 1$ and $\gamma = 1$, the estimator reduces to the standard lasso; when $\alpha = 0$ and $\gamma = 1$, the estimator reduces to the group lasso, and so on for other subsets of the three penalty terms.

\subsection{Optimization Algorithm}
\label{sec:opt}
While the optimization problem (\ref{eq:opt}) is convex for convex loss functions, due to the fusion penalty term, it is non-separable across groups of $\bm{\beta}$, so block-wise gradient descent strategies often employed for group lasso are not directly applicable. However, one algorithm that works in this case is the alternating direction method of multipliers (ADMM, \cite{boyd2011distributed}). 

For simplicity, we assume that the groups are non-overlapping and form a partition of $\bm{\beta}$, so that each coefficient belongs to exactly one group. For applying ADMM, we follow a strategy similar to that employed in \cite{huo2017integrative} and exploit the fact that $|\beta_j| = \sqrt{\beta_j^2}$ and $|\beta_j - \beta_{j-1}| = \sqrt{(\beta_j - \beta_{j-1})^2}$. Then we can reformulate the lasso and fusion $\ell_1$ penalty terms as sets of $\ell_{2,1}$ group penalties whose groups have only one member. The reformulated penalty has an overlapping group structure, with each coefficient belonging to both its own one-member group and one of the groups forming the partition of $\bm{\beta}$, and additionally we treat each absolute difference specified by the fusion matrix $\mathbf{D}$ as a group. If there are $p$ coefficients, $\mathbf{D}$ has $m$ rows, and there are $G$ groups that form a partition of $\bm{\beta}$, then the total number of effective groups is $p + m + G = N$. 

Using a least-squares loss function, we now write the objective function (\ref{eq:opt}) as 
\begin{equation}
\label{eq:reformobj}
{\arg \min}_{\bm{\beta} \in \mathbb{R}^p}   \ \  \frac{1}{2} \|\mathbf{y} - \mathbf{X} \bm{\beta}\|_2^2 + \sum_{j=1}^N \lambda_j w_j \| \mathbf{K}_j \bm{\beta} \|_2,
\end{equation}
with
\begin{equation*}
\{ \lambda_j , \mathbf{K}_j \} = \begin{cases}  \{ \lambda_1,  \mathbf{j}_j \}   & \text{if } j \in \{1, 2, \ldots, p\}  \\
                                                                         \{ \lambda_2,   \mathbf{d}_j \} & \text{if } j \in \{p+1, p+2, \ldots, p + m\} \\
                                                                         \{ \lambda_3,  \mathbf{G}_j \} & \text{if } j \in \{p + m + 1, p + m + 2, \ldots, p + m + G \},
                       \end{cases}
\end{equation*} 
where $\lambda_j \in \{ \lambda_1, \lambda_2, \lambda_3 \} $ are the regularization parameters for the lasso, fusion, and group lasso penalties, respectively; 
$w_j$ are group weights (for group lasso typically $w_j = \sqrt{p_j}$ where $p_j$ is the number of elements in group $j$); 
$ \mathbf{j}_j \in \mathbb{R}^p $ corresponds to the $j$th row of the $p \times p$ identity matrix; 
$ \mathbf{d}_j  \in \mathbb{R}^p $ corresponds to the $(j - p)$th row of the fusion matrix $\mathbf{D}$ in the three dimensional fusion penalty; 
and $ \mathbf{G}_j \in \mathbb{R}^{p_j \times p} $ is a sparse matrix where each row has a 1 at a column position corresponding to a member of group $j$. 

For ADMM, we introduce the auxiliary variables $\bm{\theta}_j = \mathbf{K}_j \bm{\beta}$. The optimization problem becomes 
\begin{align*}
\text{minimize}  \ \  &  \frac{1}{2}  \|\mathbf{y} - \mathbf{X} \bm{\beta} \|_2^2 + \sum_{j=1}^N \lambda_j w_j \| \bm{\theta}_j \|_2, \\
\text{subject to} \ \  & \bm{\theta}_j - \mathbf{K}_j  \bm{\beta} = \mathbf{0} \ \text{ for } \ j \in \{1, 2, \ldots, N \} .
\end{align*}
Let $\mathbf{K} = (\mathbf{K}_1 | \ldots | \mathbf{K}_N)$, $\bm{\theta} = (\bm{\theta}_1, \ldots, \bm{\theta}_N)^T$, and $\bm{\mu} = (\bm{\mu}_1, \ldots, \bm{\mu}_N)^T$. The augmented Lagrangian is 
\begin{equation*}
\mathcal{L}_\rho (\bm{\beta}, \bm{\theta}, \bm{\mu}) = \frac{1}{2}  \| \mathbf{y} - \mathbf{X} \bm{\beta} \|_2^2 + \sum_{j=1}^N \lambda_j w_j \| \bm{\theta}_j \|_2 +  \sum_{j=1}^N \left\{  \bm{\mu}_j^T (\bm{\theta}_j - \mathbf{K}_j \bm{\beta}) + \frac{\rho}{2} \| \bm{\theta}_j - \mathbf{K}_j \bm{\beta} \|_2^2  \right\}
\end{equation*}
where $\rho > 0$ is the step-size parameter and $\bm{\mu}_j$ are the dual variables for ADMM. After initialization of $\bm{\beta}$, $\bm{\theta}$, and $\bm{\mu}$, the update steps for ADMM consist of the following:
\begin{align*}
\bm{\beta}^{t+1} & =  {\arg \min}_{\bm{\beta} \in \mathbb{R}^p}   \ \  \mathcal{L}_\rho (\bm{\beta}, \bm{\theta}^t, \bm{\mu}^t) \\
\bm{\theta}_j^{t+1} & = {\arg \min}_{\bm{\theta}_j \in \mathbb{R}^{p_j}} \ \   \mathcal{L}_\rho (\bm{\beta}^{t+1}, \bm{\theta}, \bm{\mu}^t) \\
\bm{\mu}_j^{t+1} & = \bm{\mu}^t + \rho (\bm{\theta}^{t+1}_j - \mathbf{K}_j \bm{\beta}^{t+1} )
\end{align*}
For $\bm{\beta}^{t+1}$ and $\bm{\theta}_j^{t+1}$ updates, the corresponding $\mathcal{L}_\rho$ subgradient will equal zero at the optimal solution. Thus, the update for $\bm{\beta}$ is
\begin{align*}
\bm{\beta}^{t+1} = \left(  \mathbf{X}^T \mathbf{X} + \rho  \mathbf{K}^T \mathbf{K}  \right)^{-1}  \left( \mathbf{X}^T Y +  \mathbf{K}^T (\bm{\mu}^t + \rho \bm{\theta}^t)  \right).
\end{align*}
The subgradient with respect to $\bm{\theta}_j$ is
\begin{align*}
\frac{ \partial  \mathcal{L}_\rho }{\partial \bm{\theta}_j}  & =  \lambda_j w_j \frac{ \partial \| \bm{\theta}_j \|_2}{\partial \bm{\theta}_j }  + \bm{\mu}_j + \rho \left( \bm{\theta}_j - \mathbf{K}_j \bm{\beta} \right).
\end{align*}
In general, the subgradient of the $\ell_2$-norm $\| \mathbf{q} \|_2$ is $ \mathbf{q} / \| \mathbf{q} \|_2$ if $\mathbf{q} \ne \mathbf{0}$ and $\{ \mathbf{r}  \  |  \  \| \mathbf{r} \|_2 \le 1 \}$ if $\mathbf{q} = \mathbf{0}$. Therefore, if $\bm{\theta}_j \ne \mathbf{0}$, the condition  $ \partial  \mathcal{L}_\rho^{\theta_j} / \partial \bm{\theta}_j = \mathbf{0}$ implies that 
\begin{align*}
 \bm{\theta}_j \left( 1 + \frac{\lambda_j w_j}{\rho \| \bm{\theta}_j \|_2} \right) & = \mathbf{K}_j \bm{\beta} -  \frac{\bm{\mu}_j}{\rho} .
\end{align*}
Let $\bm{\eta}_j = \mathbf{K}_j \bm{\beta} -  \frac{\bm{\mu}_j}{\rho} $. The solution can be written in terms of the vector soft-thresholding operator $\mathcal{S}_{\kappa}(\mathbf{a}) = (1 - \kappa / \| \mathbf{a} \|_2 )_+ \mathbf{a}$, where $\mathcal{S}_{\kappa}(\mathbf{0}) = \mathbf{0}$ and $(\cdot)_+ = \max(0, \cdot)$:
\begin{align*}
 \bm{\theta}_j^{t+1} &= \mathcal{S}_{1/\rho}(\bm{\eta}_j) \\
  &= \left(1 - \frac{\lambda_j w_j}{\rho \|  \bm{\eta}_j\|_2}  \right)_+ \bm{\eta}_j .
\end{align*}

\subsubsection{Stopping Criteria}
We use the stopping criteria described in \cite{boyd2011distributed}. The algorithm terminates when the primal and dual residuals are small enough to achieve a linear combination of preselected levels of absolute $(\epsilon_{\text{abs}})$ and relative $(\epsilon_{\text{rel}})$ tolerance. Suitable values for $\epsilon_{\text{abs}}$ and $\epsilon_{\text{rel}}$ will depend on the specific application and scale of the data. Let the primal and dual residuals at iteration $t$ be denoted as $r^{t} =  \bm{\theta}^t - \mathbf{K} \bm{\beta}^{t}$ and $s^{t} = \rho \mathbf{K}^T (\bm{\theta}^{t} - \bm{\theta}^{t-1})$, respectively.  The stopping criteria are $\| r^t  \|_2 \le \epsilon_{\text{pri}}^t$ and $\| s^t  \|_2 \le \epsilon_{\text{dual}}^t$, where 
\begin{equation*}
\epsilon_{\text{pri}}^t = \sqrt{p} \, \, \epsilon_{\text{abs}} + \epsilon_{\text{rel}} \,  \max \{   \| \mathbf{K} \bm{\beta}^t \|_2  , \|  \bm{\theta}^t  \|_2    \},
\end{equation*}
\begin{equation*}
\epsilon_{\text{dual}}^t  = \sqrt{ | \bm{\theta}^t | } \, \, \epsilon_{\text{abs}} +  \epsilon_{\text{rel}} \,     \| \mathbf{K}^T \bm{\mu}^t \|_2  ,
\end{equation*}
and $  | \bm{\theta}^t | $ represents the number of elements in $\bm{\theta}^t $.

\subsubsection{Adaptive Step-size}
To accelerate the convergence of the ADMM algorithm, we implement an adaptive step-size, $\rho$, following the procedure proposed by \cite{he2000alternating} and implemented in \cite{huo2017integrative};
 \begin{equation*}
 \rho^{t+1} = 
 \begin{cases}
 \tau  \rho^t  & \text{if } \| r^t  \|_2 > \eta \| s^t  \|_2      \\
  \rho^t / \tau  & \text{if }  \| r^t  \|_2 < \eta \| s^t  \|_2 \\
 \rho^t & \text{otherwise} \\
 \end{cases},
 \end{equation*}
where $r^t$ is the primal residual and $ s^t $ is the dual residual at iteration $t$, defined above, and we set $\tau = 2$ and $\eta = 10$. This method helps to balance the primal and dual residuals so they converge to zero simultaneously.

\subsection{Adaptive Fused Sparse Group Lasso}
\cite{zou2006adaptive} showed that the lasso only exhibits consistent variable selection (i.e.,\ identifies the right subset of non-zero coefficients asymptotically) under a certain nontrivial condition, which includes an orthogonal design matrix $\mathbf{X}$ and $p = 2$ as special cases. The adaptive lasso, on the other hand, achieves consistent variable selection by differentially scaling the tuning parameter $\lambda$ for each coefficient by the factor $| \hat{\beta_j^*} |^{-\gamma}$, where $\hat{\beta_j^*}$ is a consistent estimator for $\beta_j$ such as the ordinary least squares estimator, and $\gamma > 0$  \citep{zou2006adaptive}. While the lasso biases nonzero coefficient estimates toward zero by a constant that is independent of coefficient magnitude, for the adaptive lasso the bias decreases as coefficients become large. Adaptive versions of fused lasso \citep{viallon2013adaptive} and group lasso \citep{wang2008note} have also been developed. In our application to a real neuroimaging dataset in Section \ref{sec:app}, we implement an adaptive version of fused sparse group lasso using ridge regression to obtain initial coefficient estimates, 
\begin{equation*}
\hat{\bm{\beta}}^{\text{ridge}} = {\arg \min}_{\bm{\beta} \in \mathbb{R}^p}   \ \  \frac{1}{2} \|\mathbf{y} - \mathbf{X} \bm{\beta}\|_2^2 + \lambda^{\text{ridge}} \| \bm{\beta} \|_2^2.
\end{equation*}
The weights $w_j$ introduced in Equation (\ref{eq:reformobj}) are defined as 
\begin{equation*}
\label{eq:weights}
w_j = \begin{cases}                                          \| \mathbf{j}_j \hat{\bm{\beta}}^{\text{ridge}} \|_1^{-1}   & \text{if } j \in \{1, 2, \ldots, p\}  \\
                                                                          \| \mathbf{d}_j \hat{\bm{\beta}}^{\text{ridge}} \|_1^{-1} & \text{if } j \in \{p+1, p+2, \ldots, p + m\} \\
                                                                         \| \mathbf{G}_j \hat{\bm{\beta}}^{\text{ridge}} \|_2^{-1} & \text{if } j \in \{p + m + 1, p + m + 2, \ldots, p + m + G \}.
                  \end{cases}
\end{equation*}

\section{Simulation Study} \label{sec:sim}

\subsection{Simulation Study Methods}
\begin{figure}
\begin{center}
\includegraphics[height=7in]{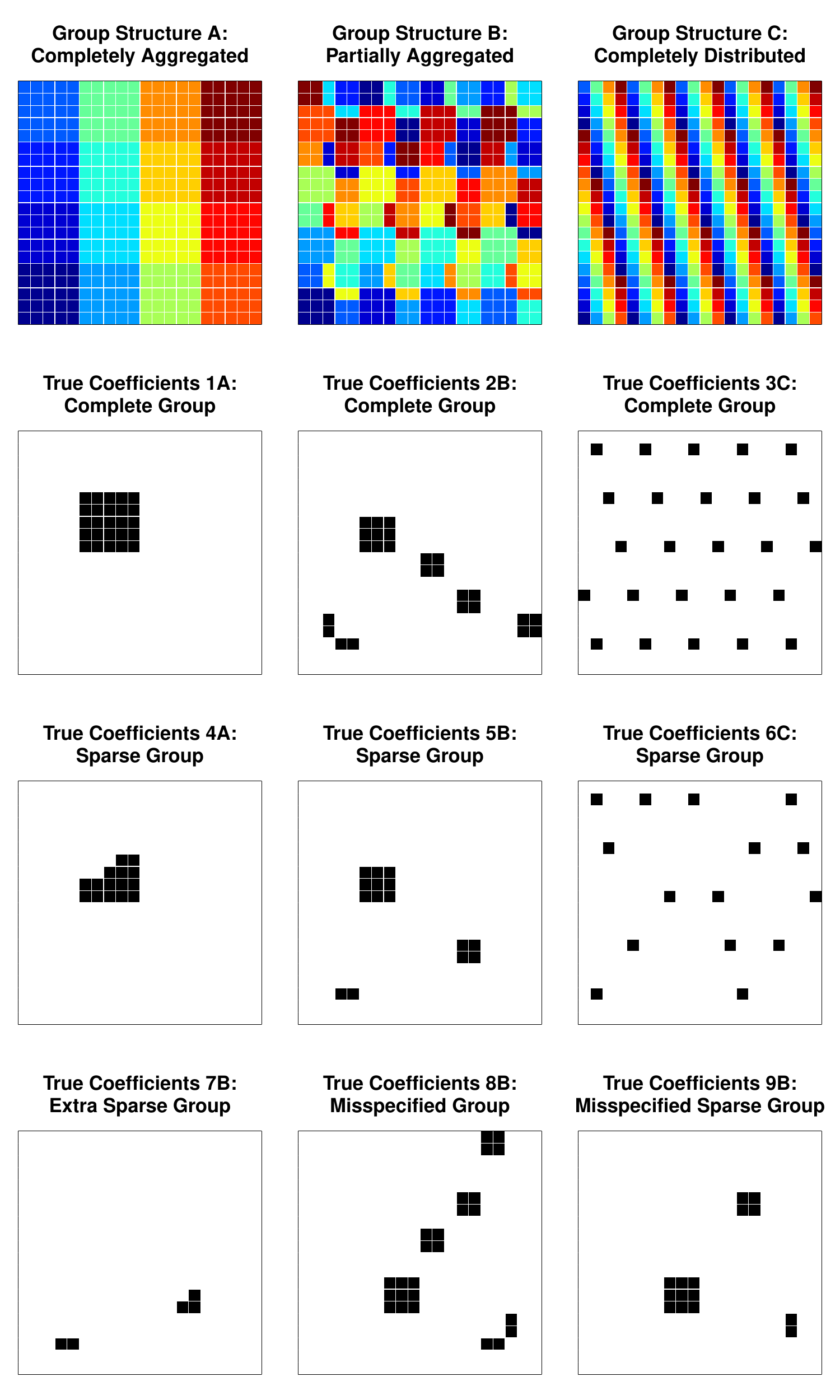}
\end{center}
\caption{Simulation study group structures (top row) and true coefficients
\label{fig:groupstructtruecoef}}
\end{figure}

This simulation study aimed to show that, for a given modeling scenario, the optimal weighting of the three penalty terms in the fused sparse group lasso depends on the underlying structure of the true coefficients, and the study aimed to characterize the optimal penalty weights for a range of different coefficient structures. Accordingly, we evenly divided the pixels of two dimensional $20 \times 20$ images into 16 groups of 25 and considered three spatial arrangements of the groups (Figure \ref{fig:groupstructtruecoef}): (A) members of a group were completely aggregated into 5 $\times$ 5 squares; (B) groups were partially aggregated, consisting of one 3 $\times$ 3 square, three 2 $\times$ 2 squares, and two 1 $\times$ 2 rectangles; (C) groups were completely distributed such that no pixels from the same group were touching sides. For each of these group structures, one group was selected to have non-zero coefficients, which were all set equal to 3. We also considered sparse versions of the coefficients, where 40\% of coefficients in the active group were set to zero. Additionally, we considered three more scenarios under the partially aggregated group structure: an extra sparse scenario, with 80\% of active group coefficients set to zero; a misspecified group structure, where the set of true coefficients was divided among several groups; and a sparse version of the misspecified group structure. Thus there were nine total scenarios of true coefficients (Figure \ref{fig:groupstructtruecoef}). 

For each of $n=50$ training subjects and $n = 50$ test subjects, we generated a vector of 400 independent standard normal random variables to serve as predictors, where each corresponded to a pixel in the $20 \times 20$ image. The responses $\mathbf{y}$ were then computed by the model $\mathbf{y} = \mathbf{X}\bm{\beta} + \bm{\epsilon}$, where each element of $\bm{\epsilon}$ was independent normal with mean zero and variance 4. To select tuning parameters, we parameterized according to Equation (\ref{eq:opt2}). For each pair of $\alpha \in \{ 0, 0.2, 0.5, 0.8, 1 \}$ and $\gamma \in \{ 0, 0.2, 0.5, 0.8, 1 \}$, we performed 5-fold cross-validation over 50 values of $\lambda = 10^x$, where values of $x$ formed a grid on the interval $[-3, 3]$, and selected the $\lambda$ that resulted in the lowest cross-validation mean squared error. We fit the model to the entire sample of $n=50$ training subjects at the given $(\alpha, \gamma)$ pair using this $\lambda$ and calculated mean squared error of the estimated coefficients $( \text{MSE}(\hat{\bm{\beta}}) = \| \hat{\bm{\beta}} - \bm{\beta} \|_2^2 / 400  )$ and mean squared prediction error for the test data $( \text{MSE}( \hat{\mathbf{y}}_{\text{test}} ) =  \| \hat{\mathbf{y}}_{\text{test}} - \mathbf{y}_{\text{test}} \|_2^2/50  )$. We repeated the entire procedure 100 times for each of the nine scenarios. We also generated another test sample of $n=100$ for the purpose of decomposing the squared bias and variance of the mean squared error at each $(\alpha, \gamma)$ combination across the 100 trained models. Analyses were done using R version 3.4.0 \citep{R2017}. 

We hypothesized that, on the basis of $\text{MSE}( \hat{\bm{\beta}} )$, $\text{MSE}( \hat{\mathbf{y}}_{\text{test}} )$, or both, (1) the fusion penalty term would perform worse and the sparsity penalty terms would perform better (i.e.,\ optimal $\gamma$ value would increase) as the groups became more spatially distributed; (2) the group penalty term would perform worse and the $\ell_1$ lasso penalty term would perform better (i.e.,\ optimal $\alpha$ value would increase) as the sparsity of true coefficients increased or with misspecification of group structure. We also sought to determine whether the lowest cross-validation error would correspond to the optimal values of $(\alpha, \gamma)$.

\subsection{Simulation Study Results}

\begin{figure}
\begin{center}
\includegraphics[width=5.5in]{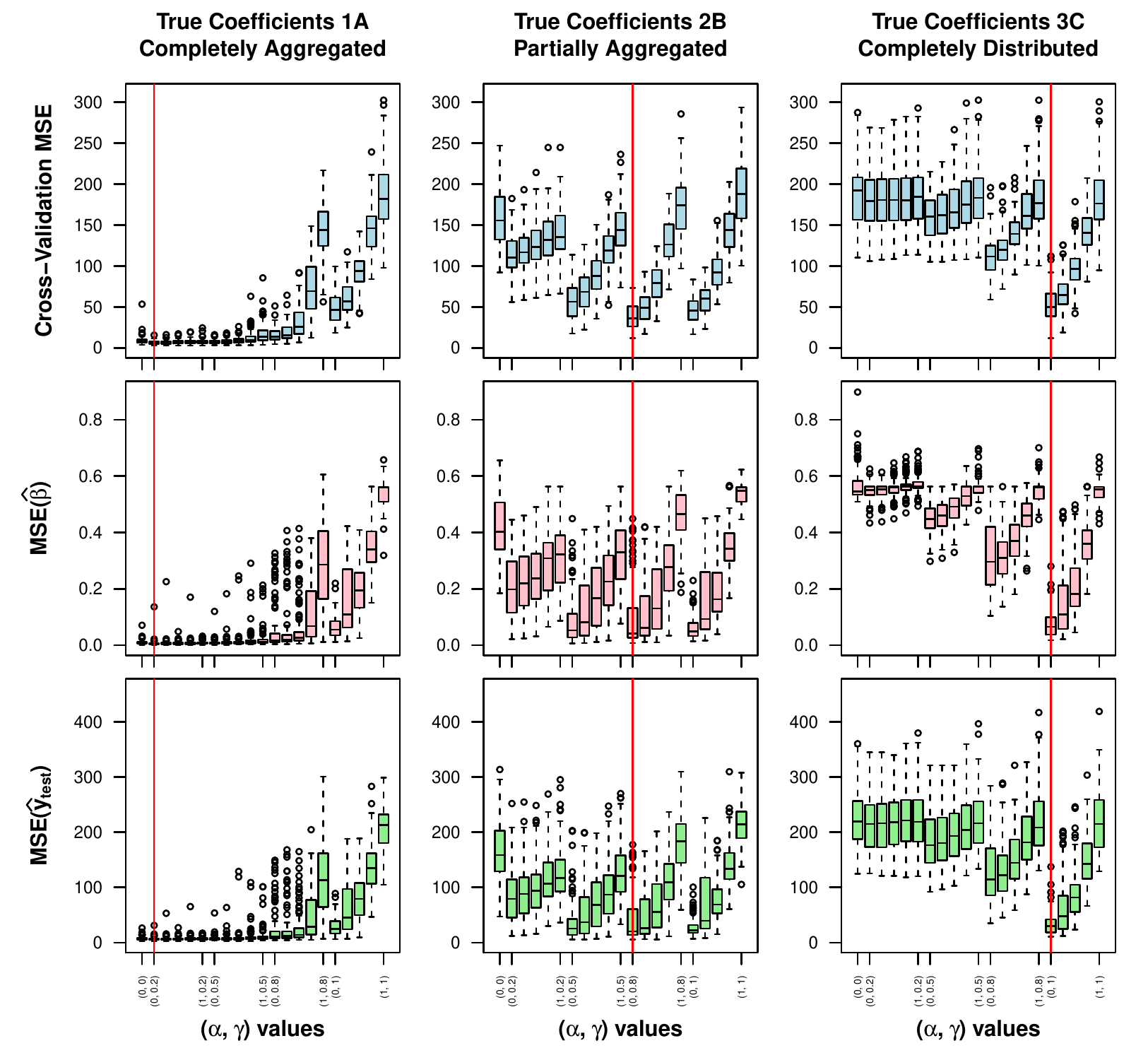} 
\end{center}
\caption[Simulation study results for true coefficients 1A, 2B, and 3C]{Simulation study results for true coefficients 1A, 2B, and 3C.  Values of $\gamma \in \{ 0, 0.2, 0.5, 0.8, 1 \}$ increase from left to right on the $x$-axis, corresponding to complete fusion penalty on the left ($\gamma = 0$) and complete sparsity penalties on the right ($\gamma = 1$). Intervals of increasing $\alpha \in \{ 0, 0.2, 0.5, 0.8, 1 \}$ values correspond to complete group penalty on the left ($\alpha = 0$) and complete $\ell_1$ lasso penalty on the right ($\alpha = 1$). Red vertical lines indicate $(\alpha, \gamma)$ combination yielding most frequent lowest error over 100 simulations. MSE: mean squared error.
\label{fig:simplot1to3}}
\end{figure}

\begin{figure}
\begin{center}
\includegraphics[width=5.5in]{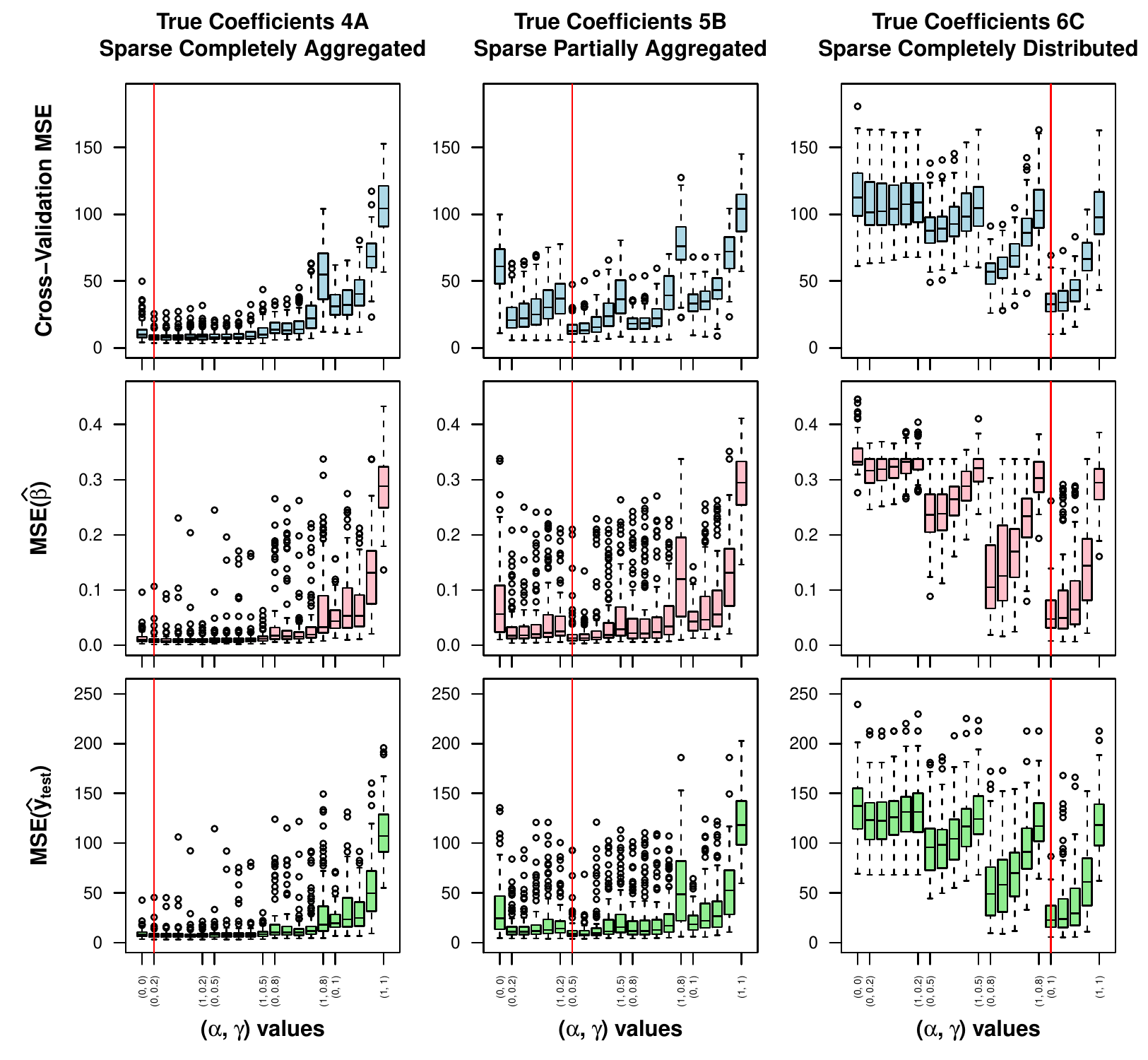} 
\end{center}
\caption[Simulation study results for true coefficients 4A, 5B, and 6C]{Simulation study results for true coefficients 4A, 5B, and 6C. Values of $\gamma \in \{ 0, 0.2, 0.5, 0.8, 1 \}$ increase from left to right on the $x$-axis, corresponding to complete fusion penalty on the left ($\gamma = 0$) and complete sparsity penalties on the right ($\gamma = 1$). Intervals of increasing $\alpha \in \{ 0, 0.2, 0.5, 0.8, 1 \}$ values correspond to complete group penalty on the left ($\alpha = 0$) and complete $\ell_1$  lasso penalty on the right ($\alpha = 1$). Red vertical lines indicate $(\alpha, \gamma)$ combination yielding most frequent lowest error over 100 simulations. MSE: mean squared error.
\label{fig:simplot4to6}}
\end{figure}

\begin{figure}
\begin{center}
\includegraphics[width=5.5in]{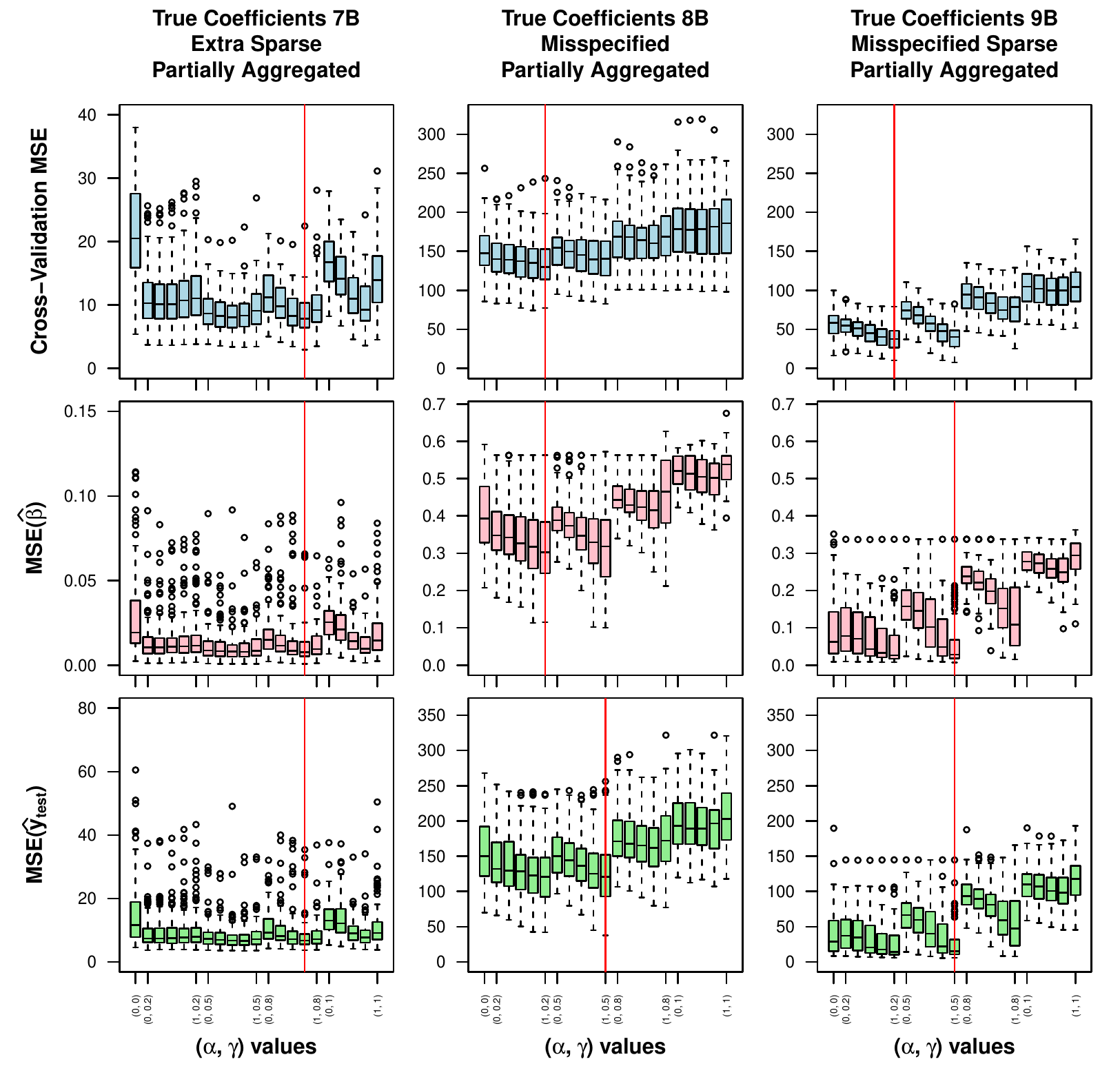} 
\end{center}
\caption[Simulation study results for true coefficients 7B, 8B, and 9B]{Simulation study results for true coefficients 7B, 8B, and 9B. Values of $\gamma \in \{ 0, 0.2, 0.5, 0.8, 1 \}$ increase from left to right on the $x$-axis, corresponding to complete fusion penalty on the left ($\gamma = 0$) and complete sparsity penalties on the right ($\gamma = 1$). Intervals of increasing $\alpha \in \{ 0, 0.2, 0.5, 0.8, 1 \}$ values correspond to complete group penalty on the left ($\alpha = 0$) and complete $\ell_1$ lasso penalty on the right ($\alpha = 1$). Red vertical lines indicate $(\alpha, \gamma)$ combination yielding most frequent lowest error over 100 simulations. MSE: mean squared error.
\label{fig:simplot7to9}}
\end{figure}

Figures \ref{fig:simplot1to3} -- \ref{fig:simplot7to9} show the distributions of cross-validation error, $\text{MSE}( \hat{\bm{\beta}} )$, and $\text{MSE}( \hat{\mathbf{y}}_{\text{test}} )$ across the $(\alpha, \gamma)$ combinations for each of the nine scenarios. Optimal $(\alpha, \gamma)$ combinations for each scenario are presented in Table \ref{tab:simresults}. Detailed simulation results are reported in Supplementary Tables S2-S10, and bias-variance decompositions of $\text{MSE}( \hat{\mathbf{y}}_{\text{test}} )$ are shown in Supplementary Figure S1.

\begin{table}
\footnotesize
\caption[Optimal ($\alpha$, $\gamma$) for each simulation scenario]{Optimal ($\alpha$, $\gamma$) for each scenario, based on the most frequent lowest error out of 100 simulation iterations \\} 
\label{tab:simresults}
\centering
\begin{tabular}{lccc}
\toprule
True coefficient scenario & Mean CVE & $\text{MSE}( \hat{\bm{\beta}} )$ & $\text{MSE}( \hat{\mathbf{y}}_{\text{test}} )$ \\ 
\midrule
1A. Completely aggregated 				& (0.0, 0.2) & (0.0, 0.2) & (0.0, 0.2) \\ 
  2B. Partially aggregated 					& (0.0, 0.8) & (0.0, 0.8) & (0.0, 0.8) \\ 
  3C. Completely distributed 				& (0.0, 1.0) & (0.0, 1.0) & (0.0, 1.0) \\ 
  4A. Sparse completely aggregated 			& (0.0, 0.2) & (0.0, 0.2) & (0.0, 0.2) \\ 
  5B. Sparse partially aggregated 			& (0.0, 0.5) & (0.0, 0.5) & (0.0, 0.5) \\ 
  6C. Sparse completely distributed 			& (0.0, 1.0) & (0.0, 1.0) & (0.0, 1.0) \\ 
  7B. Extra sparse partially aggregated 		& (0.8, 0.8) & (0.8, 0.8) & (0.8, 0.8) \\ 
  8B. Misspecified partially aggregated 		& (1.0, 0.2) & (1.0, 0.2) & (1.0, 0.5) \\ 
  9B. Misspecified sparse partially aggregated 	& (1.0, 0.2) & (1.0, 0.5) & (1.0, 0.5) \\ 
\bottomrule
\end{tabular}
\end{table}

As expected, on the basis of both $\text{MSE}( \hat{\bm{\beta}} )$ and $\text{MSE}( \hat{\mathbf{y}}_{\text{test}} )$, as groups became more spatially distributed, the optimal value of $\gamma$ increased from $0.2$ for the completely aggregated to $1$ for the completely distributed group structure, shifting weight from the fusion penalty term to the sparsity penalty terms. This pattern was similar for the complete (1A, 2B, 3C) and sparse (4A, 5B, 6C) group scenarios. As the sparsity of true coefficients increased in the partially aggregated group scenarios (2B, 5B, 7B), the optimal value of $\alpha$ increased from $0$ for the complete group to $0.8$ for the extra sparse group scenario, shifting weight from the group penalty term to the $\ell_1$ lasso penalty term. When group structure was misspecified, the optimal $\alpha$ value was 1 for both scenarios (8B, 9B), putting zero weight on the group penalty term in favor of the $\ell_1$ lasso penalty term. 

The results demonstrate that the combination of penalty terms in the fused sparse group lasso adapt to a wide range of spatial arrangements and sparsity levels of true coefficients. When group structure was correctly specified, a combination of group and fusion penalty terms was optimal for scenarios including spatially aggregated groups of true coefficients (1A, 2B, 4A, 5B), while the group penalty term alone was optimal for the completely distributed true coefficient scenarios (3C, 6C). When group structure was misspecified (8B, 9B), the fusion and $\ell_1$ penalty terms together were optimal, and for the extra sparse partially aggregated scenario (7B), all three penalty terms had non-zero weights. In all seven scenarios where group structure was correctly specified, the $(\alpha, \gamma)$ combination yielding the most frequent lowest cross-validation error corresponded to the most frequent lowest $\text{MSE}( \hat{\bm{\beta}} )$ and $\text{MSE}( \hat{\mathbf{y}}_{\text{test}} )$, indicating that selecting tuning parameters based on lowest cross-validation error tends to correspond to the optimal model. For the misspecified group scenarios, cross-validation error was lowest for either the first or second most frequent lowest $\text{MSE}( \hat{\bm{\beta}} )$ and $\text{MSE}( \hat{\mathbf{y}}_{\text{test}} )$ (see Supplementary Tables S9 and S10).

\section{Application to Neuroimaging Data}
\label{sec:app}
We applied fused sparse group lasso (FSGL) penalized regression to a resting state fMRI dataset of autism spectrum disorder (ASD, $ n=111)$ and typically developing (TD, $n=108)$ male participants (mean (SD) age 17.4 (7.5) years, see Supplementary Table S11 for descriptive summary) from the Autism Brain Imaging Data Exchange (ABIDE) repository (\cite{di2014autism}, \url{http://fcon_1000.projects.nitrc.org/indi/abide/}). In this set of participants, \cite{cerliani2015increased} used independent components analysis to identify 19 resting state brain networks. The authors found that autistic traits as measured by Social Responsiveness Scale (SRS) scores were positively associated with functional connectivity between a subcortical network, comprising basal ganglia and thalamus, and two cortical networks: (1) dorsal and (2) ventral primary somatosensory and motor cortices. The association was only significant in the ASD group. Given that the resting state networks evaluated in \cite{cerliani2015increased} represent relatively large brain regions, we used FSGL regression to more precisely define the cortical regions whose functional connectivity with a subcortical seed region best predicts SRS scores.

\subsection{Application Methods}
Preprocessed fMRI data was downloaded from the ABIDE I Preprocessed repository \citep{craddock2013neuro}. Data were preprocessed using the Connectome Computational System pipeline with no global signal regression and band pass filtering (0.01 -- 0.1 Hz) (details at \url{http://preprocessed-connectomes-project.org/abide/Pipelines.html}). The independent component resting state network data from \cite{cerliani2015increased} was downloaded from \url{https://github.com/sblnin/rsfnc} and resampled to $3 \times 3 \times 3$ mm$^3$ voxels to match the ABIDE data.  

\begin{figure}
\begin{center}
\includegraphics[width=5.5in]{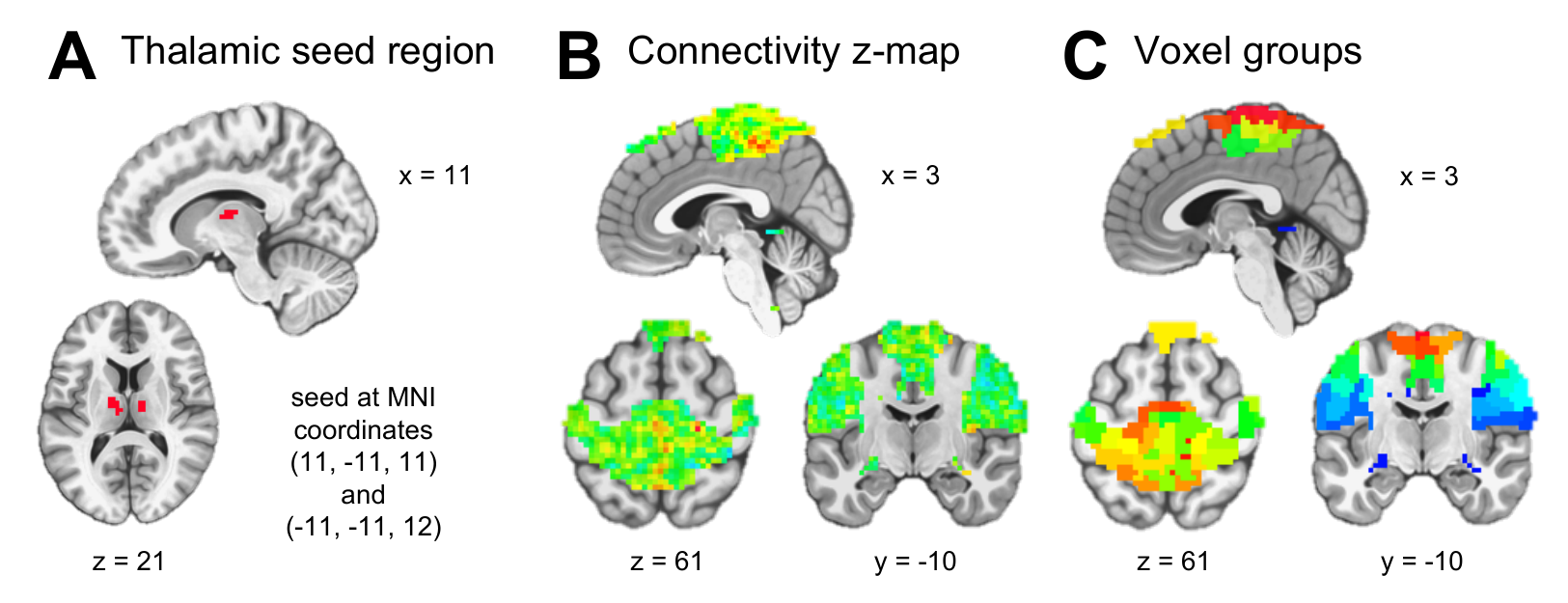} 
\end{center}
\caption{(A) The thalamic seed region consisted of 23 $3\times3 \times 3$ mm$^3$ voxels. (B) Connectivity z-maps of 5476 voxels for each participant served as predictors for fused sparse group lasso regression. (C) Voxels were partitioned into 50 groups using agglomerative hierarchical clustering on the training set resting state fMRI voxel time series. 
\label{fig:abidemethods}}
\end{figure}

We partitioned the brain into 19 resting state networks by assigning each voxel to the maximal spatial independent component at that voxel out of the 19 components identified in \cite{cerliani2015increased}. We restricted our analyses to the three networks mentioned above: the basal ganglia/thalamus subcortical network and the two sensorimotor cortical networks. We defined a subcortical seed region by selecting the peak voxels in the subcortical network independent component spatial map. This yielded bilateral regions of the thalamus, with 12 voxels centered at MNI coordinates $(11, -11, 11)$ and 11 voxels centered at $(-11, -11, 12)$ (Figure \ref{fig:abidemethods}A). For each participant, the first eigenvariate of the seed region time series was extracted, and its Pearson correlation was calculated with each voxel time series in the cortical regions of interest to form a seed-based connectivity map. Fisher's $r$-to-$z$ transformation was applied to each voxel. After excluding voxels where any participants had missing data, this left $p = 5476$ voxels to serve as predictors for FSGL regression (Figure \ref{fig:abidemethods}B).

Participant data were divided into training $(n=175)$ and test $(n=44)$ sets. Test set data was put aside until all model fitting was completed. The following steps were carried out with the training set data: 
\begin{itemize}
\item[(1)] A linear regression model adjusted raw SRS scores for age, full-scale IQ, site of acquisition, eye status at scan (open or closed), and mean framewise displacement (see Supplementary Table S12). The residuals were used as the outcome for FSGL regression.
\item[(2)] The fusion penalty was applied between coefficients of voxels who shared a face, so each voxel had a maximum of 6 neighbors. For the group penalty term, voxels in the cortical regions of interest were partitioned into 50 groups using agglomerative hierarchical clustering. First, for each training participant, Pearson correlation between time series was calculated for all possible pairs of voxels in the cortical regions of interest. Correlation matrices were averaged across participants, and a distance matrix was formed by applying the element-wise transformation $d = \sqrt{2*(1-r)}$. Finally, hierarchical clustering using Ward's method was performed based on the distance matrix, and the resulting tree was cut to form 50 groups which ranged in size from 43 to 268 voxels (Figure \ref{fig:abidemethods}C). 
\item[(3)] After standardizing columns of the predictor matrix, 5-fold cross-validation was used to determine the $\lambda$ value yielding the minimum cross-validation error at selected values of ($\alpha$, $\gamma$) for the FSGL regression. We chose to compare ($\alpha$, $\gamma$) equal to (1.0, 1.0) (standard lasso), (0.2, 1.0) (sparse group lasso), (0.2, 0.8) (fused sparse group lasso), and (0.0, 0.8) (fused group lasso). Cross-validation folds were stratified to ensure that they had similar distributions of the adjusted SRS outcome. 
\item[(4)] To estimate the coefficients, the model was fit to the entire training set at the optimal $\lambda$ for selected values of ($\alpha$, $\gamma$). 
\item[(5)] For adaptive FSGL regression, first ridge regression estimates were obtained (using R package glmnet, \cite{friedman2010glmnet}) and adaptive weights were formed according to Equation \ref{eq:weights}, and then steps (3) and (4) were completed.
\end{itemize}

For the test set data, adjusted SRS scores were predicted for each participant by taking the dot product of the estimated FSGL regression parameters with the participant's predictor variables, which were first standardized according to the training set column means and standard deviations. Raw test set SRS scores were adjusted using the linear regression model parameters estimated with the training set data. Prediction accuracy was assessed via mean squared error and Pearson correlation of predicted with actual adjusted SRS scores. Since prior studies have used resting state fMRI data to classify ASD and TD subjects into diagnostic groups rather than predict SRS, in order to compare our results we used receiver operating characteristic (ROC) curve analysis to find the best classification threshold for predicted SRS scores from the best-performing models and calculated the corresponding classification accuracies.

Analyses were done using a combination of AFNI version 17.1.03 \citep{cox1996afni}, MATLAB version 9.1.0 (R2016b) \citep{MATLAB2016}, and R version 3.4.0 \citep{R2017}.

\subsection{Application Results}
\begin{figure}
\begin{center}
\includegraphics[width=6.5in]{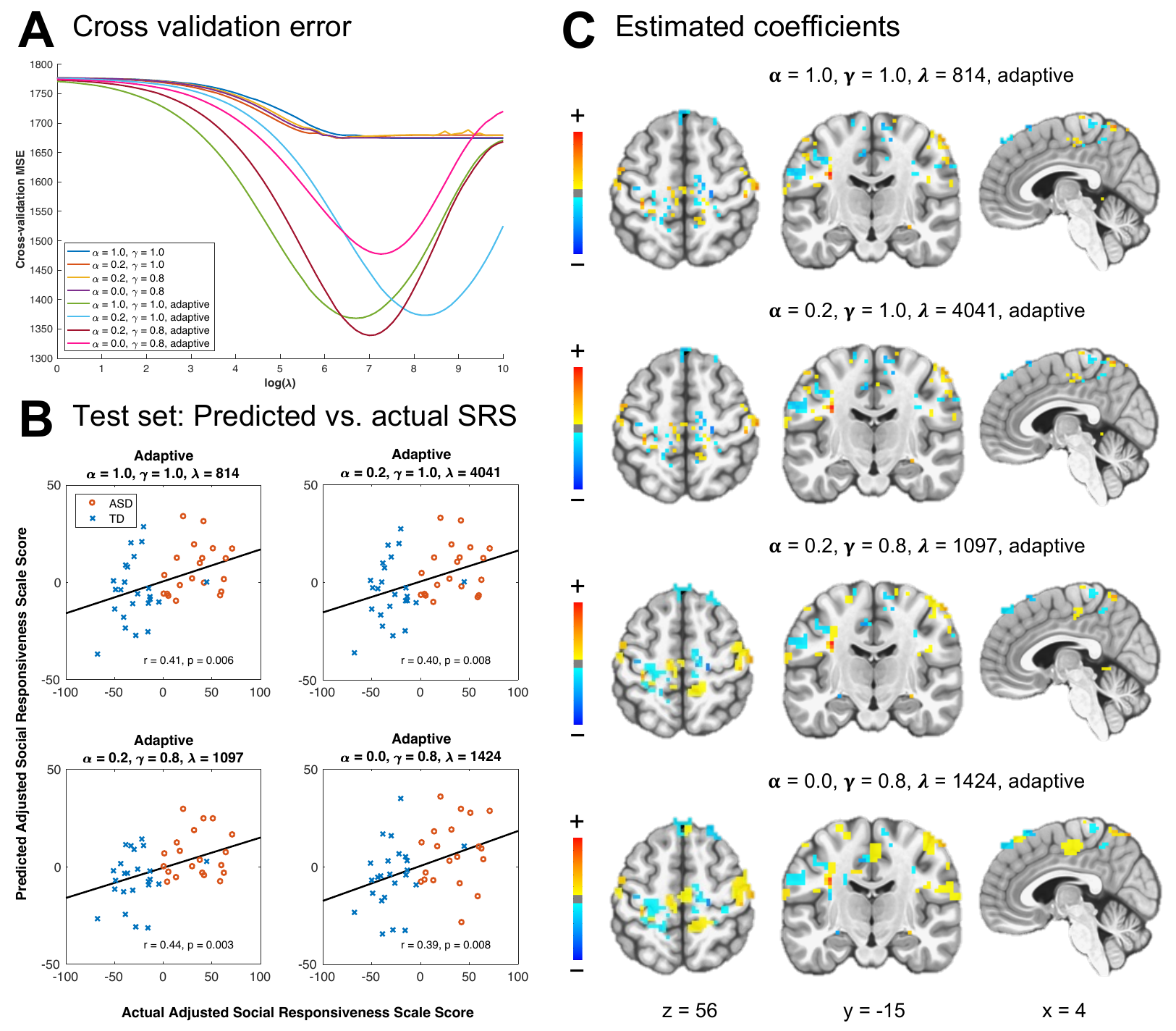} 
\end{center}
\caption{(A) Five-fold cross-validation was carried out over a range of $\lambda$ values for several sets of $\alpha$ and $\gamma$ values. (B) Correlation of predicted and actual adjusted SRS scores for selected $\alpha$ and $\gamma$ values. Points are distinguished by autism spectrum disorder (ASD, red circle) and typically developing (TD, blue cross) diagnosis groups. (C) Estimated coefficients at the optimal $\lambda$ for selected $\alpha$ and $\gamma$ values. Higher coefficient values contribute to higher predicted Social Responsiveness Scale (SRS) scores, which indicate greater autistic social impairment. 
\label{fig:abideresults}}
\end{figure}

Results for non-adaptive and adaptive fused sparse group lasso as well as for ridge and elastic net penalties are summarized in Table \ref{tab:abide}. The best test set prediction was achieved by the adaptive fused sparse group lasso with $(\alpha, \gamma)$ equal to (0.2, 0.8), which gave mean squared error of 1165.2 and Pearson correlation $r = 0.437$ $(p=0.003$) (Figure \ref{fig:abideresults}B). (For comparison, \cite{cerliani2015increased} reported correlations of $r = 0.21$ and 0.25 for the respective cortical networks in ASD participants.)  Cross-validation error was in general much lower for adaptive penalties than for non-adaptive penalties (Figure \ref{fig:abideresults}A). The superior performance of adaptive, ridge, and elastic net penalties over the non-adaptive penalties in this particular application is likely due at least in part to high multicollinearity of the predictors and the influence of many small, weak effects of predictors on the outcome rather than a few strong effects.  Estimated coefficient brain maps are shown in Figure \ref{fig:abideresults}C.  Higher SRS scores correspond to greater autistic social impairment. Thus the coefficient maps reflect multivariate thalamic seed connectivity patterns predictive of greater social impairment. Penalties including the fusion term, i.e.,\ with $\gamma = 0.8$, resulted in larger clusters of contiguous regions, rather than the more scattered coefficient maps resulting when $\gamma = 1.0$.

A couple of questions arise regarding the clinical significance of the findings. First, does the result provide insight into the neurobiology of ASD? While the sparse, structured penalty succeeded in narrowing down the predictors to a smaller subset of the most predictive voxels, it is not immediately clear why these particular scattered regions of sensorimotor cortex are most informative. We invite interested readers to further explore the coefficient brain maps available at the URL noted in the Supplementary Material section, below. Second, does the result represent a good diagnostic biomarker? We consider this question in the following context. ASD has been associated with abnormalities in connectivity between multiple brain regions \citep{di2014autism}. Accordingly, studies using resting state fMRI data to define ASD biomarkers have often summarized voxel-level data using regions of interest and considered connectivity between multiple regions, rather than use a focused, voxel-level approach as we did. Previous studies using such methods on the ABIDE dataset have achieved diagnostic classification (ASD versus TD)  accuracies in the range of 60\% to 71\% \citep{abraham2017deriving,nielsen2013multisite,plitt2015functional,kassraian2016promises}, well below the classification accuracy that can be achieved using behavioral measures such as the SRS, which can attain accuracies of up to 95\% \citep{plitt2015functional}. The difficulty of identifying fMRI biomarkers for ASD may be due to the noisiness of fMRI data or the neurobiological heterogeneity of the disorder  \citep{plitt2015functional}. What is remarkable about our result is that, when we dichotomize our predicted outcomes for the purpose of diagnostic classification, we can achieve similar accuracies (classification accuracy of 29 to 30 out of 44 test subjects, or 66\% to 68\%, for the adaptive penalties), even though we only considered connectivity between a single thalamic seed region and sensorimotor cortex. This may reflect the richness imparted by voxel-level as opposed to region of interest functional connectivity data. It seems possible that adding other features to the model, e.g., using not only a thalamic seed region, but also including voxel-level connectivity data from other seed regions that have shown abnormal connectivity in ASD, such as the default mode network and regions implicated in social cognition, would likely improve prediction performance further.

\begin{table}[]
\centering
\tiny
\caption{Comparison of estimators applied to ABIDE dataset}
\label{tab:abide}
\hspace*{-0.75cm}
\begin{tabular}{@{}llllrrrrrrrr@{}}
\toprule
					&               	&               	&                 			&         				&  \multicolumn{4}{c}{Training set $(n = 175)$}       				& \multicolumn{3}{c}{Test set $(n = 44)$}     \\   
		\cmidrule(lr){6-9}\cmidrule(lr){10-12}
Method	 				& $\alpha$ &$\gamma$ 	& Estimator 			& * Optimal $\lambda$  	&    CVMSE         	& MSE 		& $r$ 		& $p$ 		& MSE			& $r$ 		& $p$   \\ \midrule
\multirow{2}{*}{Glmnet}		& 0.0    	&     			& Ridge                             & 2627 				& 1646.7 			& 954.3		& 0.879 		&$< 0.001$ 	& 1325.8			& 0.285 		& 0.060   \\
						& 0.01 	&     			& Elastic Net                     & 289 				& 1661.2	   		& 935.7		& 0.883 		&$< 0.001$  	& 1305.4		   	& 0.320  		& 0.034     \\ \midrule
\multirow{4}{*}{FSGL}		& 1.0    	& 1.0   		& Lasso                             & 1848  				& 1674.2	   		& 1689.1	 	& 0.159	 	& 0.036	    	& 1426.7 		  	& 0.035	  	& 0.821  \\
						& 0.2    	& 1.0   		& Sparse Group Lasso      & 521  				& 1674.4	   		& 1572.0		& 0.383		&$<0.001$	& 1427.8		  	& 0.069	  	& 0.654  \\
						& 0.2  	& 0.8 		& Fused Sparse Group Lasso   & 604  			& 1673.9   		& 1633.2		& 0.254	 	&$<0.001$	& 1434.3		   	& 0.038 	  	& 0.805 \\
						& 0.0    	& 0.8 		& Fused Group Lasso       & 604  				& 1674.3	   		& 1641.2		& 0.232	 	&0.002		& 1435.4 		   	& 0.032	  	& 0.838  \\  \midrule
\multirow{4}{*}{Adaptive FSGL}	& 1.0    	& 1.0   		& Adaptive Lasso              	& 814  				& 1368.1	   		& 120.1		& 0.986 		&$< 0.001$ 	& 1193.1			& 0.406  		& 0.006    \\
						& 0.2  	& 1.0   		& Adaptive Sparse Group Lasso   & 4041      		& 1373.2        		& 129.1           	& 0.985 		&$< 0.001$      	& 1203.1               	& 0.397          	& 0.008   \\
						& 0.2  	& 0.8 		& Adaptive Fused Sparse Group Lasso & 1097    	& 1338.9	   		& 168.6		& 0.977 		&$< 0.001$ 	& 1165.2			& 0.437  		& 0.003    \\
						& 0.0    	& 0.8 		& Adaptive Fused Group Lasso    &  1424       		&  1477.2        		& 144.2           	& 0.981		&$< 0.001$	& 1211.6                	& 0.394        	& 0.008  \\ \bottomrule
\multicolumn{12}{l}{} \\
\multicolumn{12}{l}{* Note: $\lambda$ for glmnet R package is scaled by factor $n^{-1}$} \\
\multicolumn{12}{l}{Mean total sum of squares for training set = 1697.5} \\
\multicolumn{12}{l}{Mean total sum of squares for test set = 1428.0} \\
\multicolumn{12}{l}{ABIDE: Autism Brain Imaging Data Exchange} \\
\multicolumn{12}{l}{FSGL: fused sparse group lasso} \\
\multicolumn{12}{l}{CVMSE: cross-validation mean squared error} \\
\multicolumn{12}{l}{MSE: mean squared error} \\
\multicolumn{12}{l}{$r$: Pearson correlation} \\
\end{tabular}
\end{table}

\section{Conclusions}
\label{sec:conc}
The fused sparse group lasso penalty offers a flexible way to incorporate prior information into a predictive model, which can lead to more interpretable coefficient estimates and better predictive performance on test data. The fusion penalty term constrains coefficients that we expect to have similar estimated values, and we can use it to enforce local spatial smoothness in an image. The group penalty term groups together coefficients that we do not necessarily expect to have similar values, but we expect to be selected simultaneously, such as voxels residing in the same functional brain networks. The $\ell_1$ penalty term allows sparse groups, and may also be useful when groups are misspecified.  Cross-validation over a range of weights for the three penalty terms allows a data-driven way of incorporating information about coefficient structure into a prediction model. 

In this paper we have presented an ADMM optimization algorithm to fit fused sparse group lasso. A simulation study featuring a range of coefficient structures demonstrated instances where a combination of the three penalty terms together outperforms any smaller subset, and showed that cross-validation is a reliable way to select optimal tuning parameter weights. On real fMRI data, we found that incorporating adaptive weights derived from initial ridge regression coefficient estimates greatly improved performance over non-adaptive fused sparse group lasso as well as ridge and elastic net penalties. The adaptive fused sparse group lasso produced the best test set prediction, and the addition of fusion and group penalty terms resulted in less dispersed, more clustered coefficient maps. Fused sparse group lasso, a generalization of lasso, group lasso, and fused lasso, has potential application not only to prediction problems in neuroimaging, but also to other contexts where coefficients are expected to be both smooth and group-structured.

\bigskip
\begin{center}
{\large\bf SUPPLEMENTARY MATERIAL}
\end{center}

\begin{description}

\item[Supplementary Materials:] Detailed simulation study results, ABIDE dataset sample characteristics and linear regression model to adjust SRS are available in the supplementary materials. (pdf)

\item[Data and code:] Data and code for the simulation study and application to ABIDE dataset, including R and Matlab functions to fit fused sparse group lasso, are available at \url{https://github.com/jcbeer/fsgl}.

\end{description}

\bibliographystyle{agsm}
\bibliography{FSGL_Beer_et_al_2018}
\end{document}